\documentclass[lettersize,journal]{IEEEtran}
\usepackage{amsmath,amsfonts}
\usepackage{algorithmic}
\usepackage{algorithm}
\usepackage{array}
\usepackage[caption=false,font=normalsize,labelfont=sf,textfont=sf]{subfig}
\usepackage{textcomp}
\usepackage{stfloats}
\usepackage{url}
\usepackage{verbatim}
\usepackage{graphicx}
\usepackage{cite}
\usepackage{xspace}
\usepackage{enumitem}
\usepackage{tabularx}
\usepackage{multicol}
\usepackage{multirow}
\usepackage{booktabs}
\usepackage{makecell}
\usepackage{colortbl}
\definecolor{mytitle}{rgb}{0.9, 0.9, 0.9}
\definecolor{mygray}{rgb}{0.95, 0.95, 0.95}
\begin{document}

\title{RPHunter: Unveiling Rug Pull Schemes in Crypto Token via Code-and-Transaction Fusion Analysis}

\author{Hao Wu, Haijun Wang, Shangwang Li, Yin Wu, Ming Fan, Wuxia Jin, Ting Liu,~\IEEEmembership{Member,~IEEE}
        % <-this % stops a space
\thanks{Hao Wu, Haijun Wang, Shangwang Li, Yin Wu, Ming Fan, Wuxia Jin and Ting Liu are with Ministry of Education Key Laboratory for Intelligent Networks and Network Security, Xi'an Jiaotong University, Xi'an, China (e-mail: emmanuel\_wh@stu.xjtu.edu.cn; haijunwang@xjtu.edu.cn; \{3124357023, wuyin\}@stu.xjtu.edu.cn; \{mingfan, jinwuxia, tingliu\}@mail.xjtu.edu.cn)}% <-this % stops a space
% \thanks{Yitao Zhao is with Yunnan Power Grid Co., Ltd, Kunming, China (e-mail: zyt717@hotmail.com) }
\thanks{Haijun Wang is the corresponding author.}}



\maketitle

\begin{abstract}
Rug pull scams have emerged as a persistent threat to cryptocurrency, causing significant financial losses. 
A typical scenario involves scammers deploying honeypot contracts to attract investments, restricting token sales, and draining the funds, which leaves investors with worthless tokens.
Current methods either rely on predefined patterns to detect code risks or utilize statistical transaction data to train detection models. However, real-world Rug Pull schemes often involve a complex interplay between malicious code and suspicious transaction behaviors. These methods, which solely focus on one aspect, fall short in detecting such schemes effectively.

In this paper, we propose RPHunter, a novel technique that integrates code and transaction for Rug Pull detection. First, RPHunter establishes declarative rules and performs flow analysis to extract code risk information, further constructing a semantic risk code graph (SRCG). 
Meanwhile, to leverage transaction information, RPHunter formulates dynamic token transaction activities as a token flow behavior graph (TFBG) in which nodes and edges are characterized from network structure and market manipulation perspectives. Finally, RPHunter employs graph neural networks to extract complementary features from SRCG and TFBG, integrating them through an attention fusion model to enhance the detection of Rug Pull. 
We manually analyzed 645 Rug Pull incidents from code and transaction aspects and constructed a ground-truth dataset. We evaluated RPHunter on our dataset, achieving a precision of 95.3\%, a recall of 93.8\% and an F1 score of 94.5\%, which highlights superior performance compared to existing methods. Furthermore, when applied to the real-world scenarios, RPHunter has identified 4801 Rug Pull tokens, achieving a precision of 90.7\%.
\end{abstract}

\begin{IEEEkeywords}
Smart Contracts, Rug Pull, Static Analysis, Graph Neural Networks.
\end{IEEEkeywords}

\section{Introduction}
\IEEEPARstart{B}{lockchain} technology has revolutionized the financial landscape, enabling direct peer-to-peer transactions and financial services without centralized intermediaries. A diverse range of decentralized applications, including decentralized exchanges, lending platforms, and prediction markets, have been built on it. At the time of writing, the global crypto market cap has reached \$3.25 trillion~\cite{CoinMarketCap}. However, the rapid expansion and the inherent openness of DeFi also pose considerable security challenges. Among these, 
Rug Pulls have emerged as a prominent form of scam, leading to significant financial losses, with approximately \$85.4 million in 2024 alone~\cite{web3security}.

Rug Pulls are a pernicious form of scam prevalent within the DeFi ecosystem, where scammers lure investors with seemingly promising tokens. Once a substantial amount of funds is collected, these scammers abruptly withdraw the funds and abandon the project entirely, leaving investors with worthless tokens. To detect such scams, researchers have proposed various methods~\cite{ma2023pied, lin2024crpwarner, zhou2024stop, xia2021trade, mazorra2022not}, broadly classified into rule-based methods~\cite{lin2024crpwarner, zhou2024stop} and learning-based methods~\cite{xia2021trade, mazorra2022not}. 
Rule-based methods typically use static analysis to identify vulnerabilities in token contracts, requiring domain knowledge of known Rug Pull patterns. Despite their promise in detecting these known patterns, these methods are prone to high false negatives due to their limited detection scope.
In contrast, learning-based methods employ statistical features from token transactions and opcode data to train machine learning models for detection. However, these methods struggle to identify coordinated manipulation involving multiple addresses and detect Rug Pulls before they occur, suffering from high false positives and poor practicality.

The continuous evolution of real-world Rug Pulls makes them increasingly complex and diverse, posing significant challenges to existing methods. \textbf{(1)} 
Real-world Rug Pull schemes are constantly evolving and it is a complex interplay between malicious code and suspicious behaviors, yet there is still a lack of research that investigates the underlying mechanisms. Although some studies~\cite{lin2024crpwarner, zhou2024stop} have partially uncovered the mechanisms, a comprehensive empirical study focusing on the mechanisms of real-world Rug Pull cases is still missing. \textbf{(2)} 
In real-world scenarios, Rug Pulls are characterized by a complex interplay between malicious code logic and suspicious transaction behaviors. However, these two threat vectors do not always co-occur in practice. Some schemes exploit only covert code-level traps, while others rely primarily on transaction-level manipulation such as wash trading.
% In real-world scenarios, Rug Pull schemes typically begin with the injection of malicious code and then facilitate market manipulation through suspicious transaction behaviors. 
Relying solely on code or transaction information to detect Rug Pulls can lead to false negatives and false positives. However, existing methods fail to organize and leverage information from both code and transaction perspectives. 

In this paper, we focus specifically on Rug Pull schemes occurring on EVM-compatible blockchains. 
To address the aforementioned challenges, we conducted an empirical study to uncover the underlying mechanisms of Rug Pull. 
Building upon prior studies~\cite{zhou2024stop, lin2024crpwarner}, which analyzed 103 and 201 samples respectively, we construct a comprehensive real-world dataset on EVM blockchains from security platforms, comprising 645 real-world Rug Pull incidents. Each case is manually examined from both code and transaction dimensions, enabling us to summarize the common Rug Pull procedure and identify key risk features.
% We collected 645 real-world Rug Pull incidents on EVM blockchains from security platforms and manually analyzed them from code and transaction aspects.
% Compared to the most recent studies~\cite{zhou2024stop, lin2024crpwarner} with datasets of 103 and 201 samples respectively, we have constructed, to the best of our knowledge, the largest manually analyzed Rug Pull dataset and summarized the Rug Pull procedure and Rug Pull risk features. 

Building on these inherent insights, we introduce RPHunter, a novel method that models code risk information and suspicious transaction behavior within two graph structures, using graph neural networks (GNNs) for feature learning to enhance Rug Pull detection in various scenarios.
RPHunter begins by establishing declarative rules for each code risk and performing flow analysis on the bytecode to identify critical operations and relationships about code risks. These identified elements are then used to construct a semantic risk code graph (SRCG). Concurrently, RPHunter formulates dynamic token transaction activities as a token flow behavior graph (TFBG) in which nodes and edges are characterized by rich attributes from network structure and market manipulation two perspectives.
Finally, RPHunter employs GNNs to capture the features of SRCG and TFBG. By leveraging an attention fusion model, it calculates the weights of features from two graphs, integrating these features for Rug Pull detection.

We evaluated RPHunter on our ground-truth dataset. We compared the performance of RPHunter with five state-of-the-art methods. The results demonstrate that RPHunter detects Rug Pull tokens with a precision of 95.3\%, a recall of 93.8\%, and an F1 score of 94.5\%, significantly outperforming existing methods. Additionally, to assess its real-world applicability, we applied RPHunter to newly created Ethereum Mainnet crypto tokens. RPHunter identified 4801 Rug Pull tokens from 9528 tokens. We randomly sample tokens with a 99\% confidence level and an 8\% confidence interval for manual verification. The results show that RPHunter achieves an overall precision of 90.7\%.
A prototype of RPHunter, all of our datasets, and experimental results are available online~\cite{RPhunter}.
% The precision of real-world evaluation reaches 90.7\%. 

In summary, the contributions of this paper are as follows:
\begin{itemize} [leftmargin=*]
    \item We construct a comprehensive manually analyzed Rug Pull dataset, comprising 645 Rug Pull incidents, and summarize the typical Rug Pull procedure along with risk features from code and transaction perspectives.
    % \item To overcome the limitations of existing methods that fail to fully utilize contract code and transaction information, 
    \item We propose RPHunter, a novel method that integrates the complementary information from malicious code risk and suspicious transaction behavior, utilizing GNNs for graph feature learning to detect Rug Pull schemes.
    \item Experimental results show that RPHunter outperforms the current SOTA methods on our dataset with a precision of 95.3\%, a recall of 93.8\%, and an F1 score of 94.5\%. Additionally, we have identified 4801 Rug Pull tokens in real-world crypto tokens, achieving a precision of 90.7\%.
    % \item We make the source code of RPHunter, experimental data and analysis results publicly available, along with detailed Markdown files~\cite{RPhunter}.
\end{itemize}

\section{Background}
\subsection{Smart Contract and Transaction}
In this paper, we focus on blockchain networks running on the Ethereum Virtual Machine (EVM), which adopt an account-based model.
Smart contracts are self-executing programs with predefined rules and logic encoded in their code, which are typically written in a Turing-complete programming language (e.g., Solidity, Viper). To deploy smart contracts on the blockchain, they need to be compiled into bytecode and then submitted to the blockchain with transactions. 
Once deployed, smart contracts are immutable and can be accessed through public functions, serving as entry points for interactions with external owned accounts (EOAs) or other contracts.
Transactions are essential as the medium for facilitating interactions within the blockchain network, acting as the vehicle for operations such as contract creation and invocation.

\subsection{Token and Token Standard}
Tokens are a kind of cryptocurrency implemented by smart contracts which manage internal book-keeping to record the information of token holders (e.g., their addresses and shares), and support token-related activities such as token balance inquiry and token transfer. Token standards specify a unified protocol for token creation, distribution, and interaction, ensuring compatibility and functionality across various platforms. A token smart contract should implement standard interfaces and emit standard events defined by token standards, otherwise, the token cannot be purchased or sold on the blockchain. 
% Tokens are mainly classified into fungible tokens and non-fungible tokens, which follow the ERC-20 and ERC-721 standards or derived standards. 
For example, the ERC-20 standard defines 6 standard interfaces and 2 standard events for transferring tokens, retrieving account balances, and approving the use of tokens, alongside events for logging transactions. 

\subsection{Rug Pull}
Rug Pulls are a prevalent form of scam within the DeFi ecosystem, frequently executed on Decentralized Exchanges (DEXs). Unlike centralized exchanges, which match buyers and sellers via orderbook, DEXs facilitate trades through liquidity pools. 
Liquidity pools are a fundamental mechanism in DEXs where tokens are paired with reputable assets (e.g., WETH), allowing users to trade tokens directly from the pool without intermediaries. The liquidity pool dynamically adjusts token prices based on the principles of supply and demand. 

In a legitimate token launch, the process typically begins with the deployment of a token contract, followed by the creation of a liquidity pool pairing the new token with a reputable asset. The token issuer then receives liquidity provider (LP) tokens representing their share of the pool. To build trust and long-term value, responsible developers often lock these LP tokens in smart contracts (e.g., via time-locks) to prevent sudden withdrawal of liquidity. To generate revenue, developers may promote their token by associating it with unique features or utility (e.g., staking rewards), and profit through mechanisms such as transaction fees or other incentive models. In contrast, a Rug Pull is a malicious scheme where mimic the initial steps of a token launch—deploying a token contract and creating a liquidity pool—but with hidden intentions. They may embed malicious code in the contract or simply choose not to lock the initial liquidity. As the token’s price rises, the scammers exploit their unrestricted control over the liquidity pool or token contract. They swiftly exchange a large quantity of scam tokens for the paired reputable asset, effectively draining the pool. This sudden removal of liquidity leaves remaining investors holding worthless tokens with no market for resale.

% A typical Rug Pull begins with scammers deploying a scam token and creating a liquidity pool using the token paired with a reputable asset. As users purchase the token, its price will rise, creating an illusion of legitimacy and attracting more investors. Once the token price increases significantly, the scammers exploit their control over the liquidity pool or token contract by swiftly exchanging a large number of scam tokens for valuable tokens. This sudden exchange drains the liquidity, leaving investors with worthless tokens.

\section{Real-world Rug Pull Study} \label{empirical}

\subsection{Data Collection and Analysis} \label{incident}
\subsubsection{Rug Pull Incidents Collection}

Existing studies~\cite{zhou2024stop, lin2024crpwarner} have provided preliminary insights into Rug Pull schemes, analyzing 103 and 201 Rug Pull samples, respectively. To enable a more comprehensive empirical study, we significantly expanded upon these datasets by collecting Rug Pull incidents from blockchain security platforms including De.Fi~\cite{De.fi}, Peckshield~\cite{PeckShield}, Certik~\cite{Certik}, SlowMist~\cite{SlowMist} and Beosin~\cite{Beosin}.
% To conduct an empirical study on Rug Pull schemes, we extensively gathered as many Rug Pull incidents as possible from blockchain security platforms including De.Fi~\cite{De.fi}, PeckShield~\cite{PeckShield}, Certik~\cite{Certik}, SlowMist~\cite{SlowMist}, Beosin~\cite{Beosin}, BlockSec~\cite{blocksec}, RugDoc~\cite{RugDoc}, Chainalysis~\cite{Chainalysis} and Solidus Labs~\cite{Solidus}.
Ultimately, we collected a total of 1048 Rug Pull incidents that occurred before May 2024, encompassing multiple blockchain networks, including ETH~\cite{Ethereum}, BSC~\cite{BSC}, FTM~\cite{Fantom}, BASE~\cite{BASE}, Arbitrum~\cite{Arbitrum} and Polygon~\cite{Polygon}. 
% we built Rug Pull dataset from real-world incidents and conducted a manual analysis from code and transaction perspectives. This dataset can serve as a valuable resource for future research aiming to utilize program analysis to assist in Rug Pull detection.

% To conduct an empirical study on Rug Pull schemes, 

\begin{figure*}[htbp]
    \centering
    \includegraphics[width=0.8\linewidth]{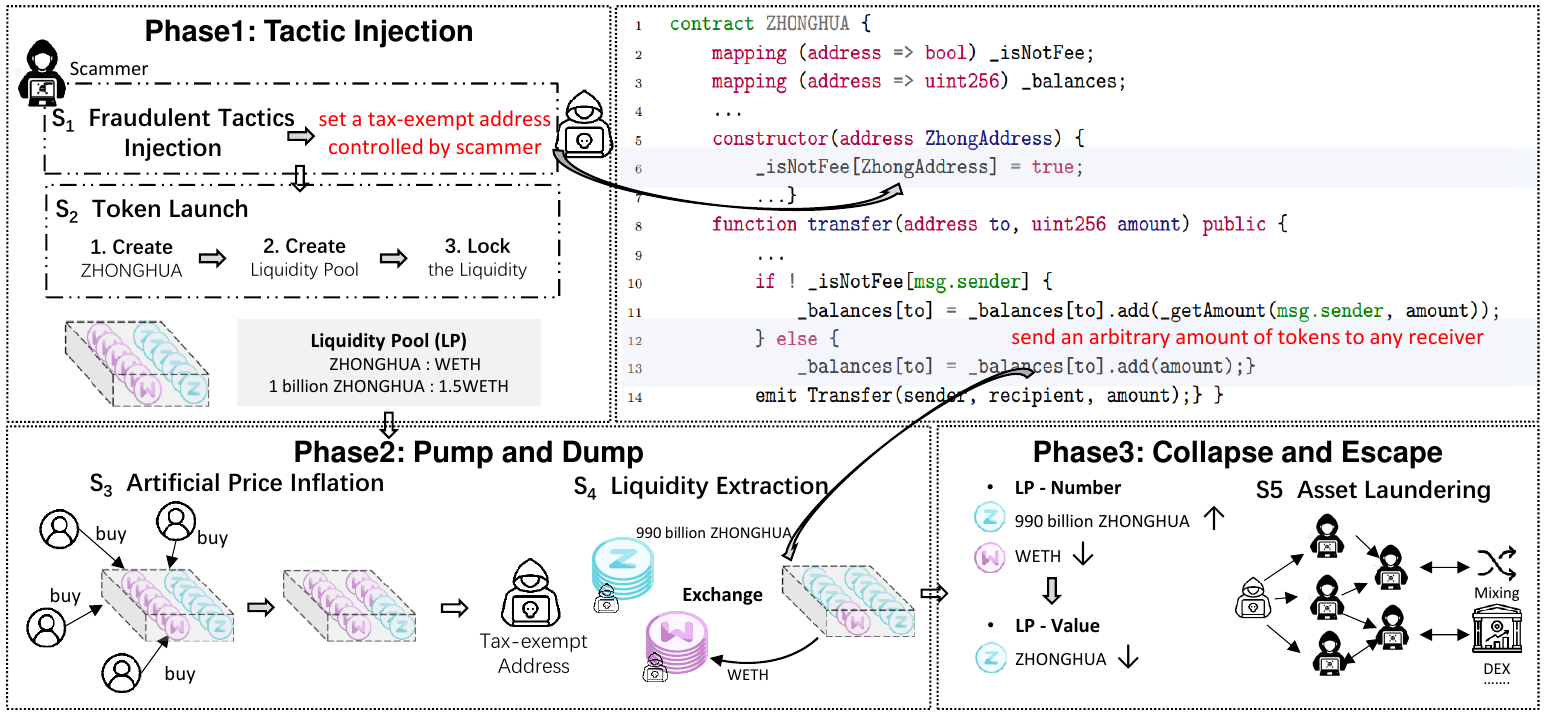}
    \caption{ZHONGHUA Rug Pull Procedure}
    \label{fig:EMPIRICAL STUDY}
    \vspace{-0.2cm}
\end{figure*}

After analyzing these Rug Pull incidents, we excluded 131 tokens without source code, hindering our precise analysis of these cases, and 272 tokens without transaction history, which suggests that these projects did not result in actual Rug Pulls. Finally, we obtained 645 Rug Pull incidents and constructed a comprehensive manually analyzed Rug Pulls dataset~\cite{RPhunter}. 
While some studies~\cite{xia2021trade, mazorra2022not} have constructed datasets of thousands of Rug Pull cases, they have two limitations: (1) these cases lack manual analysis and verification (e.g., code risk analysis), (2) rely on heuristic rules to select these cases, making them susceptible to false positives. 

\subsubsection{Incidents Analysis}
Prior studies~\cite{lin2024crpwarner,zhou2024stop, softrugpull} on Rug Pull have primarily focused on classifying these incidents based on reaction time or specific attack mechanisms. Most of these efforts emphasize the detection of malicious code pattern, often overlooking the underlying execution mechanism. 
The broader lifecycle and tactical diversity of Rug Pull remain insufficiently understood.
To better understand Rug Pull schemes, we conducted a comprehensive manual analysis of the incidents. For each Rug Pull incident, we conducted a detailed examination of the tokens, focusing on both the source code and the transaction history to uncover the underlying mechanisms of each Rug Pull. To ensure a high level of accuracy, we recruited three experts with two years of experience in smart contract security research.

Specifically, we adopted an open card sorting~\cite{spencer2009card} approach to categorize and analyze these incidents. For each Rug Pull case, we created a card comprising three components: (1) the incident name, (2) the description derived from the original incident report, and (3) the identified root cause. The root cause analysis involved two dimensions code and transaction. We performed the manual analysis, following the detailed steps illustrated in~\cite{yang2023definition}. Each expert independently examined the source code of the token contract to identify and categorize suspicious code patterns (e.g., backdoor functions). Then using CoinMarketCap~\cite{CoinMarketCap} historical data, experts reviewed each token's price trajectory to identify the precise transaction that triggered a rapid collapse in value, then analyzed the preceding transaction behaviors. After completing their independent assessments, the experts documented the identified root causes. A consensus-based voting session was then held to resolve discrepancies through discussion and determine the most plausible root cause for each incident. 
We collaboratively reviewed the identified root causes, formulating a generalized Rug Pull procedure (Section~\ref{expolit}), which captures the essence of these frauds. Additionally, we further summarized the Rug Pull risk features that emerge from code risk (Section~\ref{code risk}) and transaction behavior (Section~\ref{trading risk}) two perspectives.

% In the first round, we independently reviewed token contracts and transaction records, identifying potential code risks and unusual transaction behaviors. 
% Each expert documented the root causes of the incidents, followed by a consensus-based voting session to determine the most plausible cause for each Rug Pull. 
% The second round synthesized the insights from the initial analysis.

\begin{table*}[htbp]
\centering
\caption{Definition of Rug Pull Code Risks}
\small
\scalebox{0.9}{
\begin{tabular}{>{\arraybackslash}p{1.6cm} | >{\arraybackslash}p{4.8cm} | >{\arraybackslash}p{9.6cm}}
\toprule
\multicolumn{2}{l}{\textbf{Rug Pull Code Risks}} & 
    \textbf{Definition}
 \\
\midrule
\multirow{3}{*}{\begin{minipage}{1cm}  Sale \\ Restrict \end{minipage}} 
 & Amount Restrict (AR)  &  The volume of tokens that can be sold or held is controlled by owners \\
  & Timestamp Restrict (TR) & Sales are allowed or prohibited during specific periods controlled by owners  \\
  & Address Restrict (ADDR) & Sales are allowed or prohibited at specific addresses controlled by owners\\ 
\midrule
\multirow{3}{*}{\begin{minipage}{1cm}  Variable \\ Manipulation \end{minipage}} & Modifiable Tax Rate (MTR) & Owners can randomly change the tax rate \\
& Modifiable Tax Address (MTA) & Owners can randomly change the tax address \\
& Modifiable External Call (MEC) & Owners implement the token transfer logic in controllable external contracts\\
\midrule
\multirow{2}{*}{\begin{minipage}{1cm}  Balance \\ Tamper \end{minipage}} & Hidden Mint/Burn (HM) & Exist a hidden token mint/burn function controlled by owners \\
& Hidden Balance Modification (HBM) & Exist a hidden token balance modification function controlled by owners \\
\bottomrule
\end{tabular}}
\label{table:backdoor risks}
\vspace{-0.2cm}
\end{table*}

\subsection{Rug Pull Procedure}\label{expolit}
After a systematic analysis of the Rug Pull procedure based on these real-world incidents, we segment the procedure into three parts. We use one example to illustrate the procedure. The example is from a real-world Rug Pull incident called \textit{ZHONGHUA}~\cite{zhonghua}. Fig.~\ref{fig:EMPIRICAL STUDY} shows the simplified procedure of the scam token for illustration.

\textbf{Phase1: Tactic Injection.} 
Rug Pulls typically begin with scammers embedding fraudulent tactics to facilitate fraud ($S_{1}$: \textit{Fraudulent Tactics Injection}), followed by simulating the launch of a legitimate token, termed as $S_{2}$: \textit{Token Launch}. In $S_{1}$, scammers typically introduce hidden fraudulent tactics into the token. For instance, the token contract may contain malicious code that enables scammers to manipulate the token. Then, in $S_{2}$, scammers simulate the launch of a legitimate token. This step includes creating and promoting the new token to attract investors and establishing a liquidity pool by pairing the token with reputable assets to boost credibility. To further deceive investors, they may lock the initial liquidity, leading investors to believe that it is secure and irreversible.

\textbf{Example:} The scammer designated a tax-exempt address (line6 in Fig.~\ref{fig:EMPIRICAL STUDY}) controlled by the scammer within the token contract. Then they launched the token and locked the initial liquidity to deveive investors, promoting the token as a valuable investment.

\textbf{Phase2: Pump and Dump}. 
As the project gains visibility, scammers focus on pumping the token price to attract further investment, termed as $S_{3}$: \textit{Artificial Price Inflation}. Specifically, 
% $S_{3}$ involves aggressive marketing campaigns, often leveraging social media to generate buzz around the token. Simultaneously, 
scammers conduct high-value transactions from different addresses, creating an illusion of strong demand and liquidity. Once  a significant amount of capital is accumulated, scammers initiate the Rug Pulls, termed as $S_{4}$: \textit{Liquidity Extraction}. In $S_{4}$, scammers leverage their control over the token or liquidity pool to withdraw valuable assets. 

\textbf{Example:}  The scammer inflated the token price by 6.5 times compared to its initial value. Then the scammer exploited the malicious code injected in $S_{1}$, exchanging 990 billion tokens for 5.884 ETH from the liquidity pool. As the address controlled by the scammer had been set as the tax-exempt address, the \textit{transfer} function allows the scammer to send an arbitrary amount of tokens to any receiver bypassing balance check and token deduction (line 13).

\textbf{Phase3: Collapse and Escape}. The immediate result of a Rug Pull is a rapid devaluation of the token. As the liquidity is drained, the token price rapidly collapses to zero, making the token worthless. And scammers typically engage in laundering the stolen funds to obscure their origins, termed as $S_{5}$: \textit{Asset Laundering}. 

\textbf{Example:} after extracting liquidity, the token price immediately plummeted by a factor of 4000, leaving investors with worthless tokens. And the scammers laundered the illegal gains through various addresses, making it difficult to track back the Rug Pull.

\textbf{Summary and Insights. }
While the above procedure outlines a generalized lifecycle of typical Rug Pull schemes-begining with the injection of malicious tactics ($S_{1}$) and the new token launch($S_{2}$), followed by inflating the token price to lure investors($S_{3}$), and extracting liquidity to drain valuable assets($S_{4}$), our empirical analysis reveals that not all Rug Pull strictly adhere to this full sequence. In particular, we observed two notable variations: 
\begin{itemize} [leftmargin=*]
    \item In some cases, scammers rely solely on malicious logic embedded in the token contract, such as backdoor functions that allow unrestricted minting. They do not engage in artificial price inflation, but instead wait for organic token appreciation before silently exploiting the backdoor.
    \item Other cases show that scammers can perform Rug Pulls without introducing malicious code. Instead, they deliberately avoid locking all initial liquidity, allowing them to silently withdraw the unlocked liquidity at an opportune moment without invoking exploitative code.
\end{itemize}

These variations highlight the heterogeneity of Rug Pull schemes. A key insight from our study is that effective Rug Pull detection, especially before the scheme reaches the irreversible Phase 3, requires the integration of both code-level risk indicators from Phase 1 and suspicious transaction behaviors from Phase 2.
Focusing exclusively on one dimension may lead to potential false positives and false negatives. 
Therefore, we aim to propose a hybrid approach on EVM-compatible blockchains by jointly analyzing token contract code and early-stage transactional behaviors, enabling early-stage Rug Pull detection across heterogeneous scam patterns.
% Therefore, a hybrid detection approach that integrates code and transaction information is essential for accurately identifying diverse Rug Pull schemes in practice.

% Phase1 involves embedding malicious code that facilitates control over the token and Phase2 includes suspicious transaction behaviors that simulate organic growth. 
% Effective Rug Pull detection before reaching Phase3 relies on integrating code risk information from Phase1 and suspicious transaction behavior from Phase2, which are complementary.
% In real-world scenarios, Rug Pulls do not always exhibit both stages explicitly. 

\subsection{Rug Pull Risk Feature} \label{risk}
Based on above, we categorize the inherent characteristics of Rug Pull schemes into two types of risk features: code risk and transaction behavior. For each risk feature, we perform a detailed analysis to enhance understanding of Rug Pulls.

\subsubsection{Rug Pull Code Risk}\label{code risk}
We analyze the source code within Rug Pull tokens, categorizing code risks into three major categories and eight subcategories, expanding prior work with a more fine-grained taxonomy. Specifically, we have identified \textit{Sale Restrict} behaviors in 203 Rug Pull token contracts, \textit{Variable Manipulation} in 144 contracts, and \textit{Balance Tamper} in 189 contracts. Our analysis reveals the prevalence of these code risks in different Rug Pull schemes.
Table~\ref{table:backdoor risks} shows the definition of Rug Pull code risks.

\textbf{Category-1: Sale Restrict.} 
This category focuses on imposing restrictions on token transfers, primarily aiming to prevent investors from liquidating their holdings. 
% There are three distinct forms of sale restrictions.
First, \textit{Amount Restrict} sets constraints on the range of sale amounts, as well as minimum holding requirements, further restricting the ability of investors to sell freely. Second, \textit{Timestamp Restrict} imposes temporal limitations, prohibiting token sales during certain time windows. Third, \textit{Address Restrict} uses whitelists or blacklists to selectively permit or block certain addresses from selling tokens.

\textbf{Category-2: Variable Manipulation.} 
This category encompasses scenarios where scammers simulate legitimate behaviors while retaining control over key variables, enabling malicious manipulation. \textit{Modifiable Tax Rate} refers to cases where scammers claim that a fixed percentage of transactions will be taxed for project dividends but retain the ability to arbitrarily increase the tax rate. Similarly, \textit{Modifiable Tax Address} allows scammers to designate an address for tax collection while retaining the ability to modify it, redirecting funds to personal wallets. Additionally, under the guise of enhancing reusability, scammers implement token transfer logic in an external contract controlled by them, allowing them to replace it with malicious code, known as \textit{Modifiable External Call}.

\textbf{Category-3: Balance Tamper.} 
This category involves manipulative mechanisms that enable scammers to arbitrarily modify account balances through pre-coded functions within token contracts. It manifests in two primary forms. One exploitative form, \textit{Hidden Mint/Burn}, allows for the creation and destruction of tokens arbitrarily without transaction event records or user notifications. The other form, \textit{Hidden Balance Modification}, involves the unauthorized alteration of token balances within user accounts.

\subsubsection{Rug Pull Transaction Behavior}\label{trading risk}
We also reveal several distinct transaction behaviors that characterize these Rug Pull fraudulent schemes. These behaviors can be grouped into two overarching types: Abnormal Network Structures and Market Manipulation.

\textbf{Behavior-1: Abnormal Network Structure.}
When token transfers are visualized as a transaction graph, Rug Pulls often exhibit abnormal network structures. These networks often display centralized control points, where certain nodes exhibit unusually high connectivity and centrality, suggesting a dominant role in directing liquidity and token flows. To validate this observation, we constructed token transfer graphs for Rug Pull tokens and normal tokens in our dataset and analyzed several key structure indicators, detailed in Section~\ref{features}. We compared the distribution of these indicators across two types of graphs and observed significant differences, indicating the abnormal network structure in Rug Pulls. For instance, \textit{Indegree Centrality} and \textit{Outdegree Centrality}~\cite{wu2024tokenscout} for the nodes in the Rug Pull transfer graphs are six times higher than those in normal transfer graphs.
Thus, analyzing influential nodes and dense clusters can help identify potential Rug Pull activities.

\textbf{Behavior-2: Market Manipulation.}
Another prevalent behavior in Rug Pulls is market manipulation, aimed at creating a deceptive appearance of liquidity and trading activity. This manipulation typically involves addresses controlled by the scammers, working collaboratively to purchase large quantities of tokens within a short time period. These transactions create the illusion of high demand and rapid price growth, luring unsuspecting investors into buying the token at inflated prices. To further disguise the manipulation, scammers may distribute the transactions across various addresses to simulate trading activity, making it difficult to distinguish between legitimate price increases and artificial price manipulation. For example, we analyzed the trading activity of NexF~\cite{NexF}, a Rug Pull token, and found a remarkable symmetry between the buy and sell volumes on the Uniswap. Specifically, over a 20-day period, many addresses purchased the token and sold it afterward, inflating the trading volume and creating an illusion of increased liquidity. 

\section{METHODOLOGY} \label{methodology}

\subsection{Overview}
\begin{figure}[h]
    \vspace{-0.2cm}
    \centering
    \includegraphics[width=0.88\linewidth]{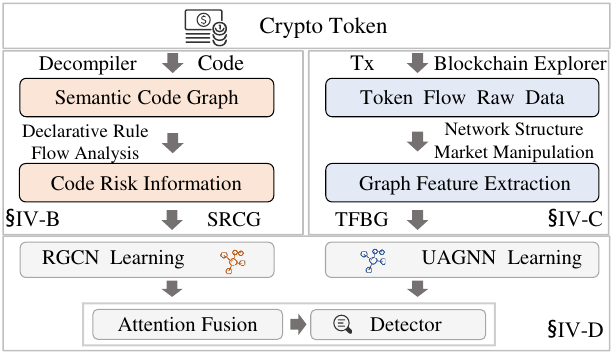}
    \caption{Overview of RPHunter}
    \label{fig:overview of RPhunter}
    \vspace{-0.4cm}
\end{figure}

Fig.~\ref{fig:overview of RPhunter} shows the overall design of RPHunter, consisting of three main modules: Semantic Risk Code Graph (SRCG) Construction, Token Flow Behavior Graph (TFBG) Construction, and Rug Pull Detection. Given a crypto token, RPHunter initially constructs the semantic code graph of the token contract, followed by establishing declarative rules for each code risk and performing flow analysis to extract code risk information. These extracted critical operations and relationships are then used to construct a semantic risk code graph. 
Concurrently, RPHunter filters all token flow transactions and leverages the raw data to extract the features of accounts and transactions based on abnormal transaction behaviors: abnormal network structure and market manipulation. These features are used to construct a token flow behavior graph. Finally, RPHunter leverages graph neural networks to represent features of SRCG and TFBG. An attention fusion model is employed to balance these features, enabling an effective detection of Rug Pulls.

% \begin{equation}
% \label{deqn_ex1a}
% x = \sum_{i=0}^{n} 2{i} Q.
% \end{equation}

\subsection{Semantic Risk Code Graph Construction}

\begin{figure*}[h]
    \centering
    \includegraphics[width=\linewidth]{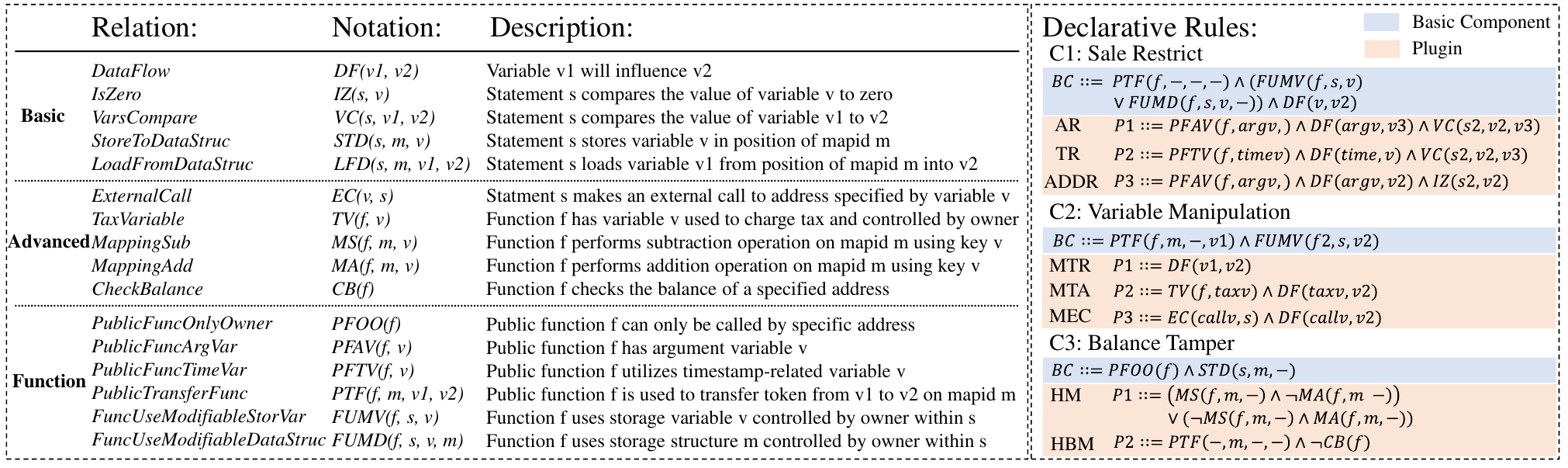}
    \caption{Relations and Declarative Rules}
    \label{fig:natation}
    \vspace{-0.2cm}
\end{figure*}

\subsubsection{Code Preprocessing}
Given a crypto token address, RPHunter firstly uses the Web3 API~\cite{Web3} \textit{getCode} to fetch the bytecode of the contract. Then it leverages the open-source binary lifting decompiler \textit{Gigahorse}~\cite{grech2019gigahorse} to decompile the contract bytecode and generate the intermediate representation (IR). The IR of the contract consists of a series of basic blocks. 
Each block encapsulates a sequence of statements. Furthermore, the control flow between two blocks is determined based on the predecessor and successor IDs. A basic block may have multiple predecessors and successors, indicating complex conditional relationships.
Based on the IR of a contract, RPHunter employs a depth-first search (DFS) method for each function to traverse and construct the control flow graph (CFG). Simultaneously, RPHunter analyzes relationships between data definitions and usage, introducing data dependency relationships. Finally, RPHunter constructs a directed semantic code graph (SCG), denoted as $SCG=(B, E)$, where $B$ is the set of nodes representing the basic blocks and $E$ is the set of edges indicating the control flow and data flow relationships between these blocks.

\subsubsection{Code Risk Information Extraction} 
To capture potential operations and relationships indicative of Rug Pull within SCG, RPHunter performs a flow analysis on the IR. Specifically, RPHunter starts by defining the relations. Based on these terms, it establishes the declarative rules for each code risk and performs flow analysis. Importantly, the objective of the flow analysis is to identify highly suspicious operations (i.e., node) and relationships (i.e., edge) within SCG that may indicate the potential presence of a Rug Pull.

\noindent \textbf{Relations. }
Here, we use \textit{s} to denote a statement, \textit{v} to denote a variable, \textit{f} to denote a function, and \textit{m} to denote a mapping structure. To better represent core operations and relationships within the code, we design some basic relations shown in Fig.~\ref{fig:natation}.
The five basic relations shown in Fig.~\ref{fig:natation} are induced with the assistance of Gigahorse. For instance, \textit{DF(v1, v2)} captures the data-flow dependency between \textit{v1} and \textit{v2}, and \textit{VC(s, v1, v2)} detects whether a comparison operation is performed between \textit{v1} and \textit{v2} within \textit{s}. 

Building upon these basic relations, we define advanced relations to better generalize semantic information in tokens. One example is \textit{TV(f, v)}, which detects the manipulated tax-related storage variable \textit{v} in \textit{f}. It identifies storage variables controlled by privileged addresses, marking them as sensitive variables. And since Solidity lacks float types, transaction tax rates are often computed using opcode \textit{MUL}, which are marked as sensitive operations. If there are reachable paths between sensitive variables and operations, it flags the variables as tax rate variables.

Furthermore, according to the summarized code risks, Rug Pull code risks predominantly occur within privileged functions and token transfer functions. Thus, we design some function relations that encapsulate the properties and operations. 
% \textit{PublicFuncOnlyOwner(func)} identifies public \textit{func} that can only be executed by specific privileged addresses. 
For example, \textit{PFAV(f, v)} recognizes the public function \textit{f} and identifies its argument variable \textit{v} the function passes. 
% \textit{PublicTransferFunc(func,mapid,var1,var2)} identifies public \textit{func} responsible for token transfer operations. It identifies \textit{func} interacting with a mapping like balance mapping identified by \textit{mapid} to verify if key-value read and write operations are performed on \textit{var1} like sender and \textit{var2} like recevier.
\textit{FUMV(f, s, v)} identifies function \textit{f} that utilizes a storage variable \textit{v} within \textit{s}, which can be modified by privileged addresses.
As the focus of this paper does not revolve around the induction of these relations, the details of their induction calculations are not furnished in this manuscript. More information can refer to Gigahorse~\cite{grech2019gigahorse} and our prototype tool~\cite{RPhunter}.

\noindent \textbf{Declarative Rules. } To effectively capture critical nodes and edges potentially indicative of code risk, we establish ``Basic Components ($BC$) + Plugins ($Pi$)'' declarative rules shown in Fig.~\ref{fig:natation}. Each of the three major categories of code risk is represented by a basic component, which captures the shared characteristics common across their respective subcategories. Tailed to the eight subcategories of code risks, specific plugins are developed to detect unique patterns specific to the subcategories, ensuring a fine-grained analysis of code risks. 

Using \textit{Address Restrict (ADDR)} as an example, the category \textit{Sale Restrict} is primarily observed within token transfer function that conceals mechanisms restricting token sales. In the basic component termed as C1-BC, we first utilize relation \textit{PTF} to identify token transfer function.
These restricted token sales based on amount, timestamp, or address are controlled by the privileged addresses, involving storage variables and mappings. Then we extract these storage variables and mappings controlled by privileged addresses, denoted as tainted data (\textit{PUMV $\lor$ PUMD}).
% Furthermore, sale restrictions based on amount, timestamp, or address are implemented by comparison operations between sensitive variables and storage variables or mappings controlled by owners. Then We extract these storage variables and mappings controlled by privileged address, which are used in the token transfer function.
Additionally, we perform flow propagation analysis on these storage variables and mappings to identify other tainted data. In \textit{ADDR}, the plugin termed as C1-P3 treats the arguments of token transfer function (i.e., address variable) as sensitive variables (\textit{PFAV}). By analyzing flow propagation, it examines whether these sensitive variables are compared against the tainted data defined in C1-BC, indicating potential constraints (\textit{VC}). Notably, due to direct comparison operators like ``$>$'' or ``$<$'' are not used in address restriction (C1-P3), the comparison operation analysis relies on the \textit{IZ(s, v)} to check whether specific conditions are compared to zero. 

\begin{figure}[h]
    \centering
    \includegraphics[width=0.99\linewidth]{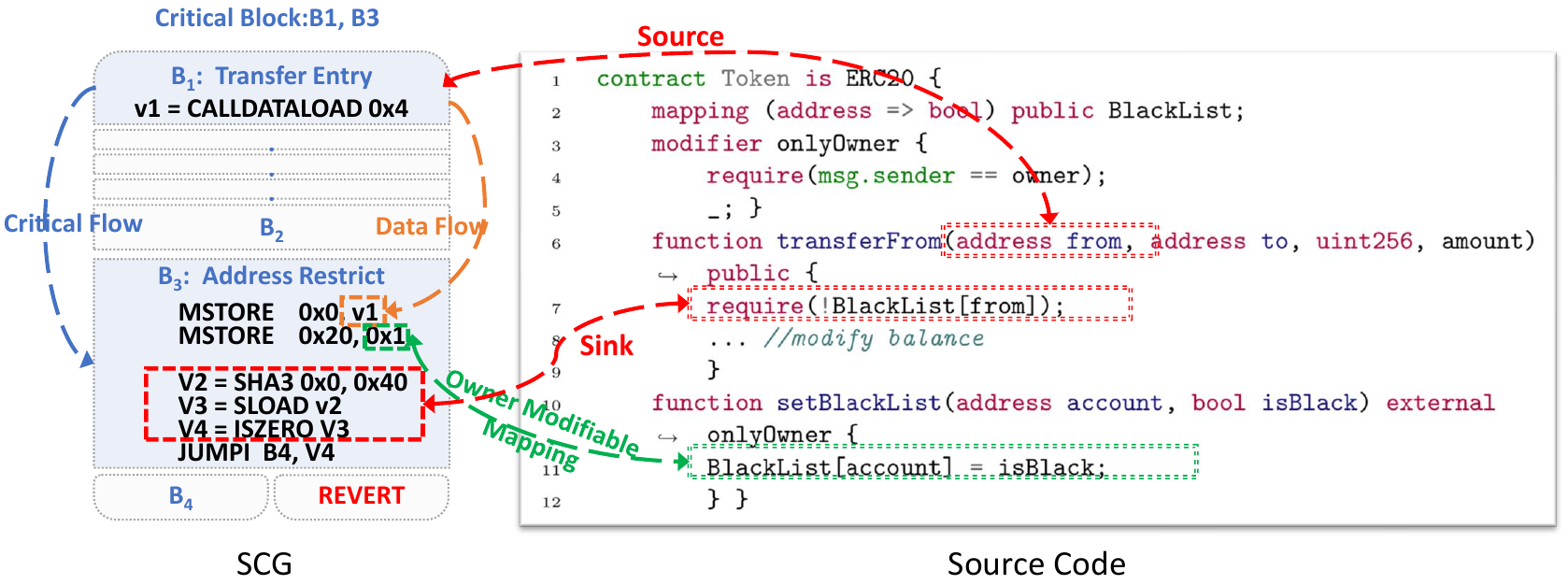}
    \caption{Example of Flow Analysis}
    \label{fig:flow analysisi example}
    \vspace{-0.15cm}
\end{figure}

\noindent \textbf{Flow Analysis.} Based on the declarative rules outlined, we propose a flow analysis framework to identify the semantics of code risk effectively. Specifically, given a SCG of the token, we combine the basic component and plugin of each code risk, denoted as $Ci:$$BC$ $\land$ $Pj$, to capture key operations and relationships. An example of \textit{Address Restrict (ADDR)} is illustrated in the Fig.~\ref{fig:flow analysisi example}. 

The \textit{transferFrom} function first checks whether the \textit{from} address is on a blacklist. If the address is blacklisted, the transaction is terminated (line 7). The blacklist is controlled by the owner (line 11), which is a typical \textit{Address Restrict} code risk. Using the declarative rule, the basic component \textit{BC} initially identifies the use of a controllable blacklist in the token transfer process (i.e., the mapId 0x1 in SCG) and marks it as tainted data.
The plugin \textit{P3} treats the \textit{from} parameters of the \textit{transferFrom} function as sensitive variables (i.e., variable v1 loaded by \textit{CALLDATALOAD} in B1 of SCG). 
% The sensitive variable is variable v1 loaded by \textit{CALLDATALOAD} in B1 of SCG. 
In B3, \textit{SHA3} computes the hash value of the content in memory from 0x0 to 0x40 (where this part involves the mapId 0x1 and the sensitive variable v1), thus v2 corresponds to the storage location of \textit{BlackList[from]}. The boolean value stored in the location is propagated to v3 and then compared against zero, forming the conditional check in \textit{require} statement (line 7). In basic blocks of SCG, there exists a reachable path between the source and sink.
Thus, we treat Blocks B1 and B3 along with their directed edges as key operations and data flows. For each code risk, the flow analysis captures the corresponding risk information involving critical operations and data flows, denoted as ``critical blocks (CB)'' and ``critical flows (CF)''.

\subsubsection{Semantic Risk Code Graph Generator}
Combining the SCG and code risk information (i.e., CB and CF), RPHunter constructs the semantic risk code graph, denoted as $\textit{SRCG} = (B, E)$. $B$ and $E$ are sets of nodes and edges. RPHunter firstly iterates through each node in the SCG and adds it to the SRCG. Then it examines each successor of the current node, and adds the successor \textit{succ} to the SRCG. During this addition, RPHunter determines the type of node (i.e., critical, invocation, normal). If the node is contained with CB, it is classified as ``critical''. If the node includes internal calls, it is labeled as ``invocation''. Otherwise, it is considered as ``normal''. Following this, RPHunter establishes the edge between the \textit{node} and its successor \textit{succ}. During this step, it determines the edge type based on their relationship. If the pair (\textit{node}, \textit{succ}) is found within CF, the edge type is classified as ``critical''. If there is a dependency relationship between the \textit{node} and the \textit{succ}, the edge type is labeled as ``dependent''. Otherwise, the edge is classified as ``normal''.

\subsection{Token Flow Behavior Graph Construction}

\subsubsection{Transaction Preprocessing} 
Based on the empirical study in Section~\ref{empirical}, it is noted that token flow can be an important indicator of suspicious transaction behaviors.
% Analyzing how tokens move between addresses provides critical insights into both code-level risks and transaction anomalies. From the code perspective, behaviors like hidden minting can manifest through unusual token flows, for instance, when an address is able to transfer out tokens despite not having received any, suggesting hidden token creation. From the transaction perspective, artificial price inflation can be indicated by observing certain addresses frequently purchasing within a short time period. 
% Consequently, our primary focus is on transactions that involve token flows.
Consequently, our primary focus is on transactions involving token flows. Given a token contract address, we leverage the blockchain explorer to collect transaction receipts involving the token flow. We focus on the earliest 500 transactions~\cite{wu2024tokenscout}, as these initial transactions can help capture suspicious transaction behavior before the occurrence of Rug Pull. A detailed discussion on the impact of the number of selected transactions is provided in Section~\ref{effectiveness}. 
For each transaction receipt, essential attributes such as the transfer value and timestamp are extracted as raw data to characterize the transaction. 
However, the variation in decimal places across different tokens can cause differences in the magnitudes of transfer values. Thus, token transfer values are processed in two steps. First, values are normalized using decimal scaling $v^{'}=v/10^{dec}$, where $v^{'}$ and $v$ denote the normalized and origin value. $dec$ denotes the number of decimal places in the smallest unit of the token. Second, a logarithmic transformation is applied to these values, helping mitigate the impact of extreme outliers.

\subsubsection{Token Flow Behavior Graph Generator} \label{features}
% Based on above transactions, we construct a token flow behavior graph (TFBG) to capture dynamic transaction activities, denoted as $\textit{TFBG} = (V, E, F_{V}, F_{E})$. $V$ is the set of nodes. $E$ is the set of edges, represented as ordered pairs $(v_{1}, v_{2})$. 
% Two nodes may be connected by multiple edges, each distinguished by unique timestamps. Each node $v \in V$ and edge $e \in E$ is associated with the feature vectors $f_{v} \in F_{V}$ and $f_{e} \in F_{E}$. 

Based on above transactions, we construct a token flow behavior graph (TFBG)  to capture dynamic transaction activities.
For each transaction, we identify the sender and receiver as nodes, connected by a direct edge. To better represent potential suspicious transaction behaviors, we enrich the TFBG by extracting a variety of features for both nodes and edges. 
Formally, we denote the TFBG as $\textit{TFBG} = (V, E, F_{V}, F_{E})$. 
% $V$ is the set of nodes, consisting of accounts involved in token flow. $E$ is the set of edges, represented as ordered pairs $(v_{1}, v_{2})$, indicating directed token transfer relationships. 
$V$ is the set of nodes. $E$ is the set of edges, represented as ordered pairs $(v_{1}, v_{2})$. 
Two nodes may be connected by multiple edges, each distinguished by unique timestamps. 
Each node $v \in V$ and edge $e \in E$ is associated with the feature vectors $f_{v} \in F_{V}$ and $f_{e} \in F_{E}$.

\begin{table*}[htbp]
\centering
\caption{Extracted Node and Edge Features in Constructing TFBG}
\small
\scalebox{0.9}{
\begin{tabular}{>{\arraybackslash}p{0.5cm} >{\arraybackslash}p{1.5cm}  >{\arraybackslash}p{4.2cm}  >{\arraybackslash}p{11.5cm}}
\toprule

% \multicolumn{2}{l}{\textbf{Notations}} & \textbf{Definition} \\
 & \textbf{Type} & \textbf{Feature} & \textbf{Description} \\
\midrule
\multirow{5}{1.5cm}{Node} & Structure & Centrality and Clustering (8) &  Capture node importance from different views \\ 

\cmidrule(l{2pt}r{2pt}){2-4}
& \multirow{4}{2cm}{Investment}
& IfTokenCreator & Check whether the node is the creator of the token contract\\
& & FundInOutRadio & The ratio of funds flowing into and out of the node \\
& & Short-term Max Incoming Tx (2) & The maximum \textbf{amounts} and \textbf{numbers} of incoming Tx over a short time period\\
& & Short-term Max Outgoing Tx (2) & The maximum \textbf{amounts} and \textbf{numbers} of outgoing Tx over a short time period\\

\bottomrule
\multirow{10}{*}{Edge} & 
\multirow{3}{*}{\begin{minipage}{1cm} \centering Time$-$series \end{minipage}} & CreationInterval & The time interval between the creation tx of the contract and the current Tx  \\ 
 & & LatestInterval & The time interval between the most recent Tx and the current Tx  \\
 & & IfApprove  & Check whether an approve tx precedes the transferFrom Tx \\
 
\cmidrule(l{2pt}r{2pt}){2-4}
% \multirow{3}{*}{\begin{minipage}{1cm} \centering Transaction\end{minipage}} 
& \multirow{3}{2cm}{Transaction} 
& Gas Limit &  The maximum gas the initiator is willing to pay for the Tx \\
& & Transfer Value &  The normalized amount of tokens transferred \\
& & Harmonic Transfer Value & The transfer value modulated by the harmonic mean of the nodes' centrality \\
\cmidrule(l{2pt}r{2pt}){2-4}
% \multirow{4}{*}{\begin{minipage}{1cm} \centering Investor \\ Tendency \end{minipage}} 
& \multirow{4}{2cm}{Investment} 
& Cumulative Incoming Tx (2) &  Total \textbf{amounts} and \textbf{numbers} of incoming Tx before the edge's timestamp\\
& & Cumulative Outcoming Tx (2) & Total \textbf{amounts} and \textbf{numbers} of outgoing Tx before the edge's timestamp \\
& & Short-term Max Incoming Tx (2) & The maximum \textbf{amounts} and \textbf{numbers} of cumulative incoming Tx over a short time period\\
& & Short-term Max Outgoing Tx (2) & The maximum \textbf{amounts} and \textbf{numbers} of cumulative outgoing Tx over a short time period\\
\bottomrule
\end{tabular}}
\label{table:node and edge features}
\end{table*}

Based on the Rug Pull transaction behaviors: Abnormal Network Structure(ANS) and Market Manipulation(MM), outlined in Section~\ref{trading risk}, we extract features from two key perspectives: network structure and manipulative behavior. The network structure features are designed to capture graph properties that may indicate ANS, while the manipulative behavior features focus on identifying outliers in specific attributes that could indicate suspicious MM.

\noindent \textbf{Node Feature Extraction.} Node features can be extracted from both network structure and manipulative behavior two perspectives. 
% Specifically, for node features within the TFBG, we refer to these as network structure and investment behavior, respectively.
The node features are shown in Table~\ref{table:node and edge features}. In terms of the network structure, we delve into the structural node attributes by calculating centrality measures and clustering coefficients, including degree centrality, indegree centrality, outdegree centrality, betweenness centrality, closeness centrality, eigenvector centrality, katz centrality, and clustering coefficient. These metrics provide insights into the node prominence and connectivity within the network, helping identify potential manipulation nodes. 

For manipulative behaviors, we define six features based on the analysis of account investment behaviors. IfTokenCreator determines whether the node is the creator of the token contract, as contract creators often have the ability to execute privileged operations. FundInOutRatio measures the ratio of funds flowing into and out of an account. An unusually high ratio across addresses may suggest a strategy to artificially inflate token prices by accumulating significant amounts of tokens. Short-term Max Incoming and Outgoing Tx measure the maximum amounts and numbers of token transfer over a short time period. A sudden spike in either incoming or outgoing transactions can indicate market manipulation.

\noindent \textbf{Edge Feature Extraction. } Edge features are extracted solely from the manipulative behavior perspective to characterize interactions between nodes.
% In the TFBG, edges represent transactions involving token flows. To adequately capture potential malicious transaction activities indicative of Rug Pulls, we leverage the transaction raw data and connected nodes to enrich the edge attributes. 
Edge features about manipulative behavior are classified into three categories: time-series, transaction, and investment, outlined in Table~\ref{table:node and edge features}. Time-series features include three indicators. CreationInterval measures the time elapsed from token contract creation to the transaction. A short interval may indicate premeditated actions immediately after contract deployment. LatestInterval measures the time interval between the current transaction and the last transaction. Short intervals in high-amount trading may suggest wash trading common in Rug Pulls. IfApprove checks whether an approve operation preceded a transferFrom transaction. Transfers lacking prior approval might indicate unauthorized token movements, raising concerns about privileged operations.

We define three features based on transactional behaviors. Gas Limit represents the maximum gas an initiator is willing to pay for a transaction. A high gas limit may suggest an urgency to complete transactions, often seen when scammers rush to complete transactions before intervention. Transfer Value reflects the normalized amount transferred. Extremely high transfer values may indicate efforts to move large amounts of tokens, manipulating market demand. Harmonic Transfer Value combines the normalized transfer value with the harmonic mean of the involved nodes' centrality, emphasizing the transactions with nodes of lower centrality and spotting networks where small transactions may conceal Rug Pulls.

To capture suspicious investment behaviors, we define eight features based on the analysis of transactions. Cumulative Incoming and Outgoing Tx quantify the total amounts and numbers of transactions where the linked node has acted as a receiver or sender up to the current timestamp. Irregularities in these metrics may signify orchestrated investment behaviors associated with scams. Similarly, Short-term Max Incoming and Outgoing Tx measure the maximum amounts and numbers of cumulative tx over a short time period, revealing abrupt token movements. Spikes in this metric may signal market manipulation attempts in a short time period, generating misleading signals to lure or panic investors. Ultimately, for each crypto token, RPHunter constructs a TFBG where both nodes and edges are characterized by 14-dimension feature vectors.

\subsection{Rug Pull Detection}\label{detection}

\subsubsection{Code Graph Embedding Learning } 
Prior work~\cite{yu2020order} has revealed the success of the Pre-train Bert model in handling program instructions. To effectively capture the semantic information, we resort to the Bert network for encoding the nodes within the SRCG. For each statement, RPHunter strips away the statement identifier and operands, retaining only the opcode. Consequently, the content of each block is distilled into a sequence of opcodes. 
% To enhance the contextual relevance of these opcode sequences, a reordering process is applied based on how blocks are linked within the SRCG. 
% By adjusting the sequences so that those from directly connected blocks are sequentially adjacent, the reordering reflects the execution flow and logical relationships. 
Subsequently, RPHunter employs the Byte Tokenizer~\cite{sennrich2015neural} for opcode tokenization and utilizes the BERT model~\cite{devlin2018bert} consisting of 12 attention heads for basic block embedding, with a dimension of 36. Finally, the node embedding within the SRCG can be computed by the trained model. Moreover, we employ a heterogeneous graph neural network to embed the SRCG. Specifically, inspired by the work~\cite{schlichtkrull2018modeling}, we implement a relational graph convolutional network (RGCN) to model the varied interactions between nodes and edges. 

The three edge types (i.e., critical, dependent, normal) in the SRCG are treated as distinct relational types within the RGCN. During the feature aggregation process, each relationship type is processed separately using a graph convolution layer where node features are updated by aggregating the features of the node and its neighboring nodes. The feature vectors from these convolutions are then aggregated across types, employing a weighted summation strategy that combines the contributions from all relationship types into a unified node feature representation as follows.

\begin{align*} \label{gcn}
   \Vec{H}^{l+1} = \sigma(\sum_{r\in R}\omega_{r}\hat{D}^{-1/2}_{r}\hat{A}_{r}\hat{D}^{-1/2}_{r}\Vec{H}^{l}W^{l}_{r})
\tag{1}
\end{align*}
% {
% \begin{equation}\label{gcn}
% \scalebox{0.95}{%
% \begin{minipage}{1.05\linewidth}
% \begin{align*}
%    \Vec{H}^{l+1} = \sigma(\sum_{r\in R}\omega_{r}\hat{D}^{-1/2}_{r}\hat{A}_{r}\hat{D}^{-1/2}_{r}\Vec{H}^{l}W^{l}_{r})
% \tag{1}
% \end{align*}
% \end{minipage}
% }
% \end{equation}}
where $R$ is the set of all relationship types, $\omega_{r}$ is the weight of relation $r$, $\hat{A}_{r}$ is the adjacency matrix for relation $r$ with added self-loops, $\hat{D}_{r}$ is the degree matrix corresponding to $\hat{A}_{r}$, $\Vec{H}^{l}$ is the node feature matrix at graph layer $l$, $W^{l}_{r}$ is the weight matrix for relation $r$ at layer $l$, and $\sigma$ is the activation function. After two layers of propagation, the node features are aggregated through mean-pooling to generate the final embedding of SRCG.

\subsubsection{Transaction Graph Embedding Learning } We use TFBG as the input graph to embed the transaction graph where both nodes and edges are represented by rich feature vectors. Instead of simply merging edge features directly into nodes, we devise Unified Aggregation GNN (UAGNN), an effective graph embedding method learning from node and edge features.
% we adopt for a more nuanced edge feature aggregation approach to fully integrate the rich information from edges. To achieve this

% To achieve this, we leverage various graph neural network architectures, including GCN , GAT, and GraphSAGE to aggregate features from both nodes and edges. After training for Rug Pull detection, we evaluate their performance to select the optimal model for the meta learner training. 

We specifically introduce the aggregation process based on GCN. In GCN, node features are updated by aggregating the features of the node itself and its neighbors, similar to Equation~\ref{gcn}. Unlike the RGCN approach, the convolution process here is more straightforward as it does not consider different edge and node types during the aggregation.
% {\setlength{\abovedisplayskip}{-1pt}
% \setlength{\belowdisplayskip}{10pt}
% \begin{equation}
% \scalebox{0.8}{%
% \begin{minipage}{1.25\linewidth}
% \begin{align*}
%    \Vec{N}^{l+1} = \sigma(\hat{D}^{-1/2}\hat{A}\hat{D}^{-1/2}\Vec{N}^{l}W^{l}_{n})
% \tag{12}
% \end{align*}
% \end{minipage}
% }
% \end{equation}}
% where $\hat{D}$ is the degree matrix, $\hat{A}=A+I_{N}$ is the adjacency matrix with added self-loops, $\Vec{N}^{l}$ is the node feature matrix at graph layer $l$, $W^{l}_{n}$ is the weight matrix, and $\sigma$ is the activation function. 
For the edge features, we not only consider the inherent features of the edge but also devise a mask mechanism to aggregate features of edges that precede the current edge in time and share the same source or target nodes as follows. 

\begin{align}
   M_{(u,v,t)} = 
   \{ e' \in S \mid\ 
   & (e'=(u, v', t')  \notag\\
   & \lor e'=(u', v, t')) \land\ (t' < t) \}
\tag{2}
\end{align}
% \begin{align*}
%    M_{(u,v,t)} = 
%    \{ e' \in S \mid (e'=(u, v', t') \lor e'=(u', v, t')) \land (t' < t) \}
% \tag{2}
% \end{align*}
% {
% \begin{equation}
% \scalebox{0.8}{%
% \begin{minipage}{1.25\linewidth}
% \begin{align*}
%    M_{(u,v,t)} = 
%    \{ e' \in S \mid (e'=(u, v', t') \lor e'=(u', v, t')) \land (t' < t) \}
% \tag{2}
% \end{align*}
% \end{minipage}
% }
% \end{equation}}
where $e=(u,v,t)$ represents an edge, $u$ and $v$ are the source and target nodes, $t$ is the timestamp of the edge, $S$ is the set of edges associated with $e$, $u'$ and $v'$ are nodes different from $u$ and $v$, $t'$ is the timestamp of the other edges, and $M_{u,v,t}$ are the relevant edges of $e$. Following that, each edge feature at graph layer $l+1$ is updated by aggregating its own features with its linked nodes and the mean value of aggregated features from other relevant edges as follows.

\begin{align*}
      \Vec{E}^{l+1}_{e} = \sigma(W^{l}_{e} \cdot (\Vec{E}^{l}_{e} \parallel \Vec{N}^{l}_{u} \parallel \Vec{N}^{l}_{v} \parallel Mean(\Vec{E}^{l}_{M_{(u,v,t)}}) ))
\tag{3}
\end{align*}
% {
% \begin{equation}
% \scalebox{0.95}{%
% \begin{minipage}{1.05\linewidth}
% \begin{align*}
%       \Vec{E}^{l+1}_{e} = \sigma(W^{l}_{e} \cdot (\Vec{E}^{l}_{e} \parallel \Vec{N}^{l}_{u} \parallel \Vec{N}^{l}_{v} \parallel Mean(\Vec{E}^{l}_{M_{(u,v,t)}}) ))
% \tag{3}
% \end{align*}
% \end{minipage}
% }
% \end{equation}}
where $W_{e}$ is the weight matrix, $\Vec{E}^{l}_{e}$ is the representation of edge $e=(u,v,t)$ at graph layer $l$, $\parallel$ is the concatenate operation, and $\Vec{N}^{l}_{u}$ and $\Vec{N}^{l}_{v}$ are the representation of nodes $u$ and $v$. After two layers of propagation, we utilize mean-pooling to aggregate the node and edge features and obtain the embedding of TFBG.

\subsubsection{Rug Pull Detection } To fully leverage the potential malicious information contained in code and transaction graphs, we utilize two graph embeddings to construct a higher level fusion model for Rug Pull detection. 
% We adopt a greedy strategy to select the optimal models for the code and transaction graphs. For the transaction graph embedding, we have utilized various GNN architectures to aggregate features, including GCN, GAT, and GraphSAGE. After training for Rug Pull detection, the results show that GraphSAGE achieves the highest F1 score (76.0\%), compared to that of GCN (74.2\%) and GAT (71.3\%). Thus, we select GraphSAGE as the base model to further construct the meta learner.
By feeding the graph embeddings produced by the two base models into the fusion model, the fusion model 
(1) maximizes the use of information from both code and transaction graphs, ensuring a comprehensive analysis of potential risks, (2) yields superior detection results by integrating the distinct but complementary insights from code and transaction graphs. 

Specifically, we perform feature alignment, mapping the graph features from SRCG and TFBG to 8-dimensional vectors to ensure consistency. Then we pass the features through the attention module, which includes sequential layers for feature transformation and weight computation. The output is a set of attention weights, which are normalized using the softmax function to obtain the final weights for both graphs. These weights are then applied to generate the overall fused feature vectors. The fused feature vectors are subsequently fed into the classifier, comprising a dropout and a fully connected layer, to obtain the predicted label for the token.

\section{EVALUATION}

We aim to address the following research questions:
\begin{itemize}[leftmargin=*]
    \item \textbf{RQ1.} How effective is RPHunter in detecting Rug Pulls compared to the SOTA methods?
    \item \textbf{RQ2.} What is the time overhead of RPHunter?
    \item \textbf{RQ3.} What are the contributions of different components of RPHunter in detecting Rug Pulls?
    \item \textbf{RQ4.} Can RPHunter unveil Rug Pull schemes in real-world crypto tokens?
\end{itemize}

\subsection{Experiment Setup}

\textbf{Dataset. }
To ensure comprehensive evaluation, we constructed our ground-truth dataset based on two principles: (1) the availability of open-source code, which facilitates reviewing and analyzing the token contract, and (2) the presence of transaction history, which allows for the analysis of transaction behaviors. We collected a total of 1048 Rug Pull tokens that occurred before May 2024. 
After excluding 131 tokens without source code and 272 tokens without transaction history, we obtained 645 real-world Rug Pull tokens, detailed in Section~\ref{incident}. Additionally, to train our model for Rug Pull detection, we collected 1806 benign tokens from an open-source dataset~\cite{wu2024tokenscout}. These tokens are real-world crypto tokens, which have already been manually inspected and labeled as benign tokens. We excluded 131 tokens from the benign tokens due to the lack of transactions. As a result, we obtained a ground-truth dataset with 645 Rug Pull tokens and 1675 benign tokens for our experiments. And to collect token flow transactions, we leveraged blockchain explorer API to get a list of Token Transfer Events by token address.

\textbf{Implementation. } We implemented the prototype of RPHunter in Python and Datalog. To extract code risk information, we defined declarative rules using Datalog based on Gigahorse~\cite{grech2019gigahorse}. We implemented Rug Pull Detector using over 2000 lines of Python, with the graph neural network designed using Pytorch~\cite{paszke2019pytorch}. The training, validation and test sets were split in a 60\%-20\%:-20\% ratio. To ensure the training set represents all samples, we used five-fold cross-validation and reported the average results.

\textbf{Evaluation Metrics.} For all experiments, we employed five metrics including Precision, Recall, F1 score, false positive rates (FPR), and false negative rates (FNR) to evaluate their performance.

All experiments were conducted on machines equipped with the Intel(R) Xeon(R) Platinum 8163 CPU @ 2.50GHz, 3*RTX3090 GPU, and 1TB RAM running 64-bit Ubuntu 18.04.6 LTS.

\subsection{RQ1: Effectiveness of RPHunter}  \label{effectiveness}
To evaluate the effectiveness of RPHunter, we compared its performance with five state-of-the-art Rug Pull detection methods. There are two selection criteria for these SOTA methods: (1) open-source availability or ease of replication, and (2) suitability for automated large-scale experiments. The five methods are Ma et al.~\cite{ma2023pied}, Lin et al.~\cite{lin2024crpwarner}, Mazorra et al.~\cite{mazorra2022not} and Xia et al.~\cite{xia2021trade} from academia and GoPlus~\cite{GoPlus} from industry. 
Ma et al.~\cite{ma2023pied} proposed Pied-Piper, which integrates datalog analysis and fuzzing to detect some Rug Pulls. Lin et al.~\cite{lin2024crpwarner} proposed CRPWaner, the most recent academic tool that detects Contract-related Rug Pulls based on static analysis. We used these two methods as the baselines for rule-based methods which are open-sourced~\cite{CRPWARNER, pied}. Mazorra et al.~\cite{mazorra2022not} and Xia et al.~\cite{xia2021trade} utilized statistical transaction features to train scam token detection models. We used these two methods as the baseline for learning-based methods, replicated their models based on their paper, and conducted experiments on our ground-truth dataset. GoPlus~\cite{GoPlus} is a widely known commercial tool, providing APIs to detect scam tokens. We used GoPlus as the baseline for commercial methods.

\begin{table}[htbp]
\centering
  \caption{Performance comparison with SOTA methods}
  \label{tab:rq1}
  \renewcommand{\arraystretch}{1.2}
  \scalebox{0.9}{
  \begin{tabular}{c|ccccc}
    \hline
    \textbf{Methods} & \textbf{Precision} & \textbf{Recall} & \textbf{FPR} & \textbf{FNR} & \textbf{F1}\\
    \hline
    Pied-Piper~\cite{ma2023pied} & 58.8\% & 5.8\% & \textbf{1.7\%} & 94.2\% & 10.5\% \\
    CRPWarner~\cite{lin2024crpwarner} & 31.2\% & 56.2\%  & 47.3\% & 43.8\% & 40.1\% \\
    Xia et al.~\cite{xia2021trade} & 70.8\% & 75.8\% & 12.1\% & 24.2\% & 73.2\% \\
    Mazorra et al.~\cite{mazorra2022not} & 71.9\% & \textbf{79.4\%} & 11.9\% & \textbf{20.6\%} & 75.5\% 
    \\
    GoPlus~\cite{GoPlus} & \textbf{74.0\%} & 78.2\% & 10.6\% & 21.4\% & \textbf{76.2\%} \\
    \textbf{RPHunter} & \textbf{95.3\%} & \textbf{93.8\%} & \textbf{1.8\%} & \textbf{6.2\%} & \textbf{94.5\%} \\
  \hline
\end{tabular}}
\vspace{-0.2cm}
\end{table}

The results can be seen in Table~\ref{tab:rq1}, where RPHunter outperforms all baselines on the ground-truth dataset, achieving a precision of 95.3\%, a recall of 93.8\%, an FPR of 1.8\%, an FNR of 6.2\%, and an overall F1 score of 94.5\%. However, Pied-Piper achieves the lowest FPR, since it only targets a small part of Rug Pull code risks. Due to limited detection capability, it naturally exhibits the lowest FPR at the cost of a extremely high FNR.
% its recall is extremely low, reaching only 5.8\%, which inherently contributes to the lowest FPR.
% We present the results in Table~\ref{tab:rq1}. As can be seen, RPhunter outperforms all the baselines on the ground-truth dataset, achieving an overall F1 score of 94.5\%, an FPR of 1.8\%, and an FNR of 6.2\%. 
% Compared to the best performance baseline, RPhunter has improved the overall detection precision, recall, and F1 score by 21.3\%, 14.4\%, and 18.3\%, respectively. 

\noindent \textbf{Impact of Code: }In the dataset, there are 229 incidents that exhibit no obvious code risks and primarily initiate Rug Pulls through unlocked liquidity, which are common~\cite{tokenchaos}. This led to relatively poor performance by the static analysis tools, Pied-Piper and CRPWarner. In contrast, RPHunter not only captures semantic information from the code but also leverages early-stage transactions that reveal malicious behaviors. This combination enables effective detection of Rug Pulls, even for tokens without apparent code risks. 

\noindent \textbf{Impact of Transaction:} Learning-based methods, due to the reliance on large transaction volumes and the absence of code analysis, also performed poorly. To further evaluate the impact of transaction volume, we analyzed the transaction distribution within our Rug Pull dataset. Our findings indicate that 84$\%$ of Rug Pull tokens have more than 100 transactions, whereas only 3.4$\%$ of tokens have fewer than 10 transactions. In fact, Rug Pull tokens typically have high transaction volumes to create the illusion of active trading and ensure sufficient profits before the scam, which are available for transaction analysis.
% This aligns with our expectations, as tokens with extremely low transaction volumes often fail to gain investment before executing a Rug Pull. 
Next, we conducted controlled experiments under two conditions: with and without code information, denoted as \textit{Tool} and \textit{$Tool_{NoC}$}. Specifically, we tested the performance of RPHunter at transaction volumes of 100, 200, 300, 400, and 500, denoted as \textit{$Tx_{num}$}. The results are presented in Table~\ref{tab:rq1-2}. 

\begin{table}[htbp]
  \caption{Impact of Transaction Volume on Performance}
  \label{tab:rq1-2}
  \renewcommand{\arraystretch}{1.2}
  \scalebox{0.84}{
  \begin{tabular}{c|ccc|c|ccc}
    \hline
     \textbf{$Tool_{NoC}$}  &  \textbf{Precision} &  \textbf{Recall} & \textbf{F1} & $Tool$ & \textbf{Precision} &  \textbf{Recall} & \textbf{F1} \\
    \hline
    % V1 & $CG_{NoSR}$ & $\times$ & 86.6\% & 81.7\% & 4.7\% & 18.3\% & 84.1\% \\
    $Tx_{100}$ & 63.6\% & 66.1\% & 64.8\% & $Tx_{100}$ & 92.1\% & 91.4\% & 91.8\% \\
    $Tx_{200}$ & 70.9\% & 68.0\% & 69.4\% & $Tx_{200}$ & 92.2\% & 93.0\% & 92.6\%  \\
    $Tx_{300}$ & 72.7\% & 75.0\% & 73.8\% & $Tx_{300}$ & 95.1\% & 91.4\% & 93.2\%  \\
    $Tx_{400}$ & 77.2\% & 74.2\% & 75.7\% & $Tx_{400}$ & 95.2\% & 93.0\% & 94.1\%  \\
    $Tx_{500}$ & 71.7\% & 80.7\% & 76.0\% & $Tx_{500}$ & 95.3\% &  93.8\% & 94.5\%  \\
  \hline
\end{tabular}}
\end{table}

The results show that when code information is not combined ($Tool_{NoC}$), the detection effectiveness gradually improves and stabilizes as the transaction volume increases. $Tx_{500}$ has improved the overall detection F1 score by 11.2\% compared to $Tx_{100}$, indicating a strong dependency on transaction volume. However, when code information is incorporated ($Tool$), the results show that even with a lower transaction volume 100, the F1 score reaches 91.8\%, and the difference between $Tx_{100}$ and $Tx_{500}$ is only 2.7\%, showing the robustness of our method. This suggests that by combining the code and transaction information, RPHunter can maintain stable and effective detection capabilities even in low-transaction scenarios.

\subsection{RQ2: Time Overhead of RPHunter}
\begin{figure}[htbp]
    \centering
    \begin{minipage}{0.47\linewidth}  % 每个minipage占据单栏的48%
        \centering
        \includegraphics[width=\linewidth]{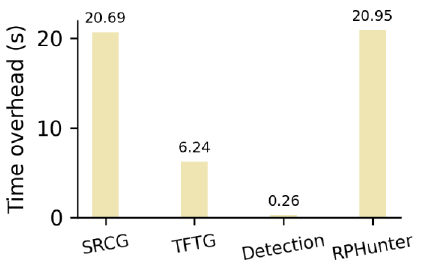}
        \caption{Overhead of Each Component}
        \label{fig:rq2-1}
    \end{minipage}%
    \hspace{0.04\linewidth}
    \begin{minipage}{0.47\linewidth}
        \centering
        \includegraphics[width=\linewidth]{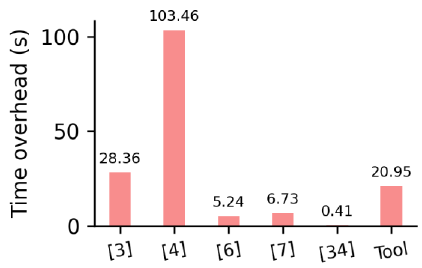}
        \caption{Overhead Comparison with SOTA methods}
        \label{fig:rq2-2}  
    \end{minipage}
\end{figure}

Given a crypto token address, we evaluated the time overhead of RPHunter for detecting Rug Pull risks within the token. To accurately evaluate the time overhead, we tested each crypto token five times in our dataset and recorded the average time. As shown in Fig.~\ref{fig:rq2-1}, the average time consumption results of each component of RPHunter is 20.69s for SRCG Construction, 6.24s for TFBG Construction, and 0.26s for Rug Pull Detection. Due to the parallel processing capabilities in constructing SRCG and TFBG, the average time overhead of RPHunter is 20.95s. This time overhead is acceptable for early Rug Pull detection, enabling proactive measures before financial losses occur. Most of the overhead of RPHunter comes from complex flow analysis.

We also calculate the time overhead for other SOTA methods on the same dataset. As shown in Fig.~\ref{fig:rq2-2}, it is evident that rule-based methods~\cite{ma2023pied, lin2024crpwarner} are time-consuming, taking 28.36s and 103.46s respectively for each crypto token. Alternatively, learning-based methods~\cite{xia2021trade, mazorra2022not}, which utilize machine learning, demonstrate a shorter detection time. The majority of its time overhead is spent on processing transaction data to extract statistical features. Learning-based methods are generally faster than rule-based methods, but lack code analysis. Furthermore, the commercial tool GoPlus~\cite{GoPlus} shows the shortest detection time in standard scenarios. However, when analyzing newly deployed tokens, GoPlus may take over 5 minutes to produce a detection result. In summary, RPHunter presents a balanced approach with an acceptable time overhead and offers a comprehensive analysis of code semantic information.

\subsection{RQ3: Ablation Study} \label{ablation study}
As described earlier, RPHunter comprises three main design components. First, it focuses on extracting code risk information during the SRCG construction. Second, it emphasizes the integration of node and edge features during the TFBG construction. Finally, RPHunter utilizes a greedy strategy to select optimal models from the SRCG and TFBG and performs a fusion analysis of the two graphs, constructing a fusion model for Rug Pull detection. To evaluate the effectiveness of these design components, we conducted an ablation study as follows.

\begin{table}[htbp]
  \caption{Ablation Study Results}
  \label{tab:rq3}
  \renewcommand{\arraystretch}{1.2}
  \resizebox{\linewidth}{!}{
  \begin{tabular}{c|cc|ccccc}
    \hline
     & \textbf{Code Graph}  & \textbf{Tx Graph} & \textbf{Precision} &  \textbf{Recall} & \textbf{FPR} & \textbf{FNR} & \textbf{F1} \\
    \hline
    % V1 & $CG_{NoSR}$ & $\times$ & 86.6\% & 81.7\% & 4.7\% & 18.3\% & 84.1\% \\
    V1 & $SRCG$ & $\times$ & 92.7\% & 89.1\% & 2.7\% & 10.9\% & 90.8\% \\
    V2 & $SRCG_{NoSR}$ & $\times$ & 86.6\% & 81.7\% & 4.7\% & 18.3\% & 84.1\%  \\
    V3 & $\times$ & $TFBG$ & 71.7\% & 80.7\% & 12.6\% & 19.2\% & 76.0\% \\
    V4 & $\times$ & $TFBG_{NoN}$ & 66.4\% & 63.1\% & 10.5\% & 36.9\% & 66.4\% \\
    V5 & $\times$ & $TFBG_{NoE}$ & 72.5\% & 58.7\% & 9.4\% & 41.3\% & 64.8\% \\
    \textbf{Tool} & $SRCG$ & $TFBG$ &  \textbf{95.3\%} & \textbf{93.8\%} & \textbf{1.8\%} & \textbf{6.2\%} & \textbf{94.5\%} \\
  \hline
\end{tabular}}
\end{table}

% \begin{table}[htbp]
%   \caption{Ablation Study Results}
%   \label{tab:rq3}
%   \renewcommand{\arraystretch}{1.3}
%   \scalebox{0.8}{
%   \begin{tabular}{c|cc|ccccc}
%     \hline
%      & \textbf{Code Graph}  & \textbf{Tx Graph} & \textbf{Precision} &  \textbf{Recall} & \textbf{FPR} & \textbf{FNR} & \textbf{F1} \\
%     \hline
%     % V1 & $CG_{NoSR}$ & $\times$ & 86.6\% & 81.7\% & 4.7\% & 18.3\% & 84.1\% \\
%     V1 & $SRCG$ & $\times$ & 92.7\% & 89.1\% & 2.7\% & 10.9\% & 90.8\% \\
%     V2 & $SRCG_{NoSR}$ & $\times$ & 86.6\% & 81.7\% & 4.7\% & 18.3\% & 84.1\%  \\
%     V3 & $\times$ & $TFBG$ & 71.7\% & 80.7\% & 12.6\% & 19.2\% & 76.0\% \\
%     V4 & $\times$ & $TFBG_{NoN}$ & 66.4\% & 63.1\% & 10.5\% & 36.9\% & 66.4\% \\
%     V5 & $\times$ & $TFBG_{NoE}$ & 72.5\% & 58.7\% & 9.4\% & 41.3\% & 64.8\% \\
%     \textbf{Tool} & $SRCG$ & $TFBG$ &  \textbf{95.3\%} & \textbf{93.8\%} & \textbf{1.8\%} & \textbf{6.2\%} & \textbf{94.5\%} \\
%   \hline
% \end{tabular}}
% \end{table}

% \subsubsection{Effectiveness of transaction number}

\subsubsection{Effectiveness of extracting code risk information in SRCG} 
We conducted a variant experiment, denoted as $SRCG_{NoSR}$ (V2), which excludes the code risk information from the SRCG. As shown in Table~\ref{tab:rq3}, compared to the $SRCG$ (V1), the precision, recall, and F1 score of $SRCG_{NoSR}$ decrease by 6.1\%, 7.4\%, and 6.7\%, respectively. The results highlight the $SRCG$ holds more fine-grained code risk semantic information in the graphical features and underscore the critical role of code risk information in improving detection performance within the SRCG. Notably, the $SRCG$ (v1) achieves a high recall of 89.1\%, even though our earlier empirical study identifies 229 Rug Pulls that exhibit no explicit code risks. Our analysis suggests that while these contracts may lack clearly malicious code, they often exhibit subtle structural differences from normal contracts. The SRCG captures these fine-grained patterns, and the Bert-based encoder effectively learns these deviations during training. As a result, the high recall of V1 stems not only from detecting known risky patterns, but also from recognizing latent structural and semantic deviations that associated with Rug Pull contracts.

\subsubsection{Effectiveness of incorporating node and edge features in TFBG} 
We conducted two variant experiments by respectively removing node and edge features from the TFBG, denoted as $TFBG_{NoN}$ (V4) and $TFBG_{NoE}$ (V5). As shown in Table~\ref{tab:rq3}, compared with the $TFBG$ (V3) with node and edge features, the absence of node features in $TFBG_{NoN}$ leads to decreases in precision, recall, and F1 score by 5.3\%, 17.6\%, and 9.6\% respectively. 
Similarly, the removal of edge features in $TFBG_{NoE}$ results in a drop of 22.0\% in recall, and 11.2\% in F1 score. The absence of node and edge features leads to a significant decrease in recall and an increase in the FNR, indicating that both node and edge features play an important role in capturing transaction behaviors. We did not conduct an experiment that removed both node and edge features, which strip all meaningful information of TFBG, rendering it ineffective for analysis. 

\subsubsection{Effectiveness of fusion analysis of SRCG and TFBG} We compared the individual performances of variant $SRCG$ (V1) and $TFBG$ (V3) against their fusion use within RPHunter. Specifically, compared to $SRCG$ alone, RPHunter shows improvements in precision, recall, and F1 score by 2.6\%, 4.7\%, and 3.7\%. The improvements indicate $TFBG$ can help RPHunter utilize transaction behavior information for detection, which cannot be captured by $SRCG$ alone. Furthermore, compared to $TFBG$ alone, the precision, recall, and F1 score of RPHunter increase by 23.6\%, 13.1\%, and 18.5\%, demonstrating that RPHunter effectively incorporates information from code patterns and provides valuable insights into code risks. This result highlights that the fusion analysis of $SRCG$ and $TFBG$ synergistically enhances the performance of Rug Pull detection.

\subsection{RQ4: Rug Pull in Real World} To evaluate RPHunter's performance on crypto tokens beyond our benchmark, we conducted preliminary experiments on the real-world Ethereum Mainnet. We collected and analyzed contracts deployed from May 1 2024 to May 29 2024, spanning blocks 19771560 to 19972630. Each contract's bytecode was examined to determine if the public function signatures conform to the ERC20 token standards. Utilizing the methodologies described in Section~\ref{methodology}, we constructed the respective SRCG and TFBG for each token and fed them into the trained model for Rug Pull detection. 

Ultimately, RPHunter successfully identified 4801 Rug Pull tokens from 9528 ERC20 Tokens. This proportion is alarmingly high, yet consistent with prior industry findings, for example, blockchain security firm Certik~\cite{tokenchaos} has reported that one in every two new tokens on Ethereum is involved in scam.
We sampled 247 open-source tokens based on a 99\% confidence level and an 8\% confidence interval~\cite{Condifence} for in-depth analysis to substantiate the detection results. This involved a thorough investigation of each contract's code patterns and transaction behaviors. Specifically, we first retrieve the source code from blockchain explorer and perform double-check manual auditing to identify the presence of predefined code risks. If any such risks were confirmed, we flagged the token as a potential Rug Pull. Then we further analyzed each token's price trajectory using CoinMarketCup~\cite{CoinMarketCap} and associated transaction records. We investigated whether the token experienced a sudden and irreversible price crash, and if such a crash was tied to suspicious liquidity withdrawals or concentrated token transfers, thereby corroborating the presence of a Rug Pull event.

Upon analysis, we found 23 false positives, and the overall true positive rate achieves 90.7\%. Through a deeper investigation into false positive cases, we identified the main reason. In certain instances, scammers refrain from writing complex contracts or initiating additional transactions to attract investment, which is costly and cumbersome, increasing the risk of detection. Instead, they just leverage off-chain social media to attract investors and exit by removing liquidity. This liquidity issue is closely tied to off-chain trust mechanisms. We plan to explore this in the future. In addition, further analysis of these Rug Pull tokens reveals that 108 of these tokens existed for no more than 24 hours, indicating that Rug Pulls are executed with remarkable speed. The massive issuance of short-lived Rug Pull token not only highlights the prevalence of fraudulent schemes but also emphasizes the necessity of the detection framework.
% highlighting the severe security risks associated with newly issued tokens on the blockchain. 
% It is worth noticing that closed-source contracts, which could not be reviewed, are considered to have a higher likelihood of being associated with Rug Pull schemes. 
We discuss one case study found on real-world Ethereum Mainnet for illustration~\cite{MNHAToken}.

\begin{figure}[htbp]
    \centering
    \includegraphics[width=\linewidth]{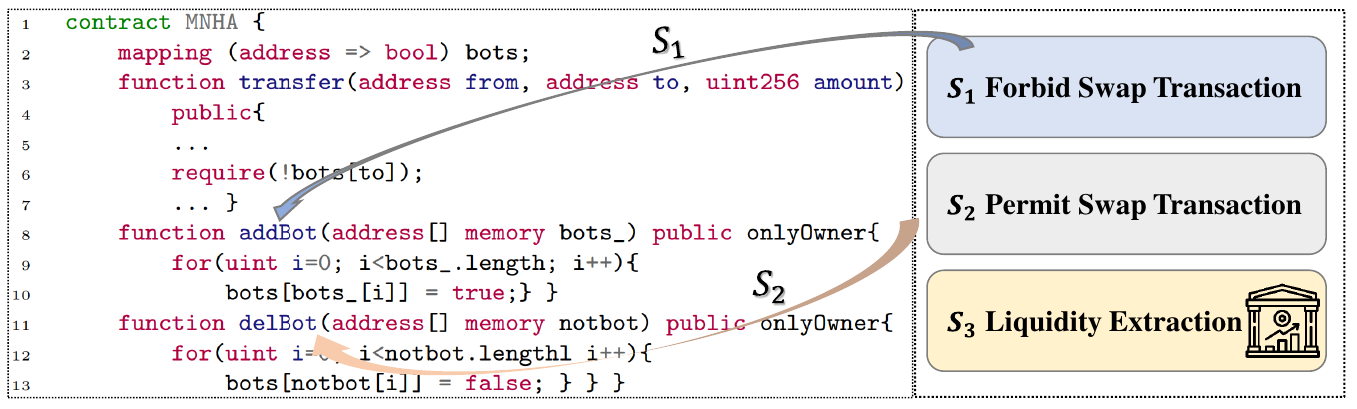}
    \caption{Case Study}
    \label{fig:rq4}
\end{figure}

\noindent \textbf{Case study on MNHA token.} Fig.~\ref{fig:rq4} shows the Rug Pull procedure. As the token price increased, the scammers triggered the \textit{Address Restrict} vulnerability (line 10). This tactic restricted the liquidity pool contract from processing transactions, effectively preventing users from swapping their \textit{MNHA} tokens ($S_{1}$). Once the scammers permitted swap transaction (line 13, $S_{2}$), they quickly utilized unlocked liquidity to exchange WETH, draining the liquidity pool ($S_{3}$). The entire process from token deployment to the completion of the Rug Pull lasted only 8 minutes. RPHunter effectively identifies the Rug Pull before the liquidity extraction occurred.

\subsection{Threats to Validity} 
\noindent \textbf{Internal Threats.}
 The effectiveness of RPHunter is threatened by the insufficiently precise flow analysis (produced by Gigahorse~\cite{grech2019gigahorse}). Since our method only leverages Gigahorse to extract code risk information, this threat can be mitigated by integrating more advanced tools. Additionally, during the real-world validation on the Ethereum Mainnet, we rely on manual inspection to determine whether a contract constitutes a Rug Pull. This subjective process may introduce labeling inconsistencies or human error. To mitigate this threat, we validated token data through trusted platform CoinMarketCap and performed a double-check mechanism to ensure accuracy.
 
\noindent \textbf{External Threats.} Our ground-truth dataset is constructed by manually collecting and analyzing a set of Rug Pull incidents. Despite our efforts to gather a comprehensive set of known cases, the process may still introduce sample bias or contain inaccurate labeling. This could affect the effectiveness of RPHunter. To mitigate this threat, we sourced incidents from multiple channels and recruited a panel of three domain experts to review and verify each case, thereby increasing the reliability of datasets.

\section{RELATED WORK}

\noindent \textbf{Rug Pull Detection.}
Scam activities are pervasive in the blockchain ecosystem, posing serious threats to the credibility and security of decentralized systems. Various types of scams, such as Ponzi schemes, and phishing attacks, have been widely studied in recent years~\cite{sun2020early, chen2021sadponzi, torres2019art, roy2024unveiling, lin2021phishpedia, xia2020don, beres2021blockchain, bartoletti2020dissecting, liebau2019crypto, sureshbhai2020karuna, wu2020phishers, liang2024ponziguard, zhou2024stop, lin2024crpwarner}. 
% For Ponzi schemes, Chen et al.~\cite{chen2021sadponzi} extracted semantic information through symbolic execution and compares it to Ponzi scheme patterns for detection. Additionally, many studies~\cite{chen2018detecting, jung2019data, fan2020expose, lou2020ponzi} utilize machine learning methods to identify Ponzi schemes. For phishing attacks, ~\cite{torres2019art, roy2024unveiling, lin2021phishpedia} identify phishing risks before users interact with phishers by scanning phishing websites and malicious smart contracts. And ~\cite{reid2013analysis, xia2020don, beres2021blockchain} focus on transaction graph analysis to identify phishing transaction.
Recently, Rug Pull has emerged as one of the most severe and damaging forms of scams~\cite{cernera2023token,huang2023deep, sechting2024taxonomy}. Some research~\cite{xia2021trade, mazorra2022not} explored machine learning techniques to identify Rug Pull tokens, utilizing transactions and statistical features. Ma et al.~\cite{ma2023pied} defined five distinct types of backdoors in smart contracts and employed a combination of datalog analysis and fuzzing techniques to detect these backdoors. Furthermore, Lin et al.~\cite{lin2024crpwarner} and Zhou et al.~\cite{zhou2024stop} performed an empirical study on Rug Pull events and exposed Rug Pull risks utilizing domain-specific datalog analysis. Commercial tools like GoPlus~\cite{GoPlus}, TokenSniffer~\cite{TokenSniffer} audit token contracts based on the source code or EVM bytecode of smart contracts. 
In contrast to existing methods that focus on either code patterns or transaction behaviors. RPHunter integrates code and transaction information for comprehensive Rug Pull detection.
% However, existing methods struggle to detect Rug Pulls in various scenarios, focusing on either risk code patterns or suspicious transaction behaviors. RPHunter integrates code and transaction information for comprehensive Rug Pull detection. 

% However, Rug Pull scams typically arise from a combination of code backdoors and malicious transaction behaviors. In contract to existing methods that focus exclusively on either code pattern or transaction history, RPHunter combine both dimensions into a unifies detection framework.

\noindent \textbf{Neural Networks for Vulnerability Detection.}
Various tools have been proposed to detect smart contract vulnerabilities using deep neural networks~\cite{wu2020comprehensive,wang2020contractward,wu2021peculiar,hu2021transaction, gopali2022vulnerability, zhuang2021smart, cheng2024vulnerability, jeon2021smartcondetect, wang2024defiguard, sendner2023smarter}. Jeon et al.~\cite{jeon2021smartcondetect} extracted code fragments from smart contracts and detected vulnerable code patterns using a pre-trained BERT model. Nevertheless, these sequence-based methods assume that the smart contract is a sequence of tokens without considering the graph structure of the contracts~\cite{huang2021hunting,zeng2022ethergis}. 
% Some studies~\cite{cheng2024vulnerability, liu2021combining,zhuang2021smart}  have compared the performance of graph-based methods with sequence-based methods, revealing the graph-based methods often yield superior results for contract vulnerability detection. 
% Graph neural networks are a class of neural networks that are designed to process and learn from graph data~\cite{wu2020comprehensive}. 
Graph neural networks have been proven to excel in vulnerability detection by leveraging the graph structure of code, which represents both the control flow and data dependency among different instructions~\cite{liu2021combining, cheng2024vulnerability}. Zhuang et al.~\cite{zhuang2021smart} and Liu et al.~\cite{liu2021combining} constructed code semantic graphs from source code and proposed TMP~\cite{zhuang2021smart} and AME~\cite{liu2021combining} to detect vulnerabilities, respectively. SCVHunter~\cite{luo2024scvhunter} designed a heterogeneous semantic graph based on intermediate representations and proposed a heterogeneous graph attention network for detection. PonziGuard~\cite{liang2024ponziguard} collected the contract's runtime behavior using fuzzing techniques and established the contract runtime behavior graph to accurately detect Ponzi schemes. 

\section{CONCLUSION}
In this paper, we propose RPHunter, a novel Rug Pull detection framework that integrates code and transaction information, modeling these information within SRCG and TFBG for feature learning to unveil Rug Pull schemes. We evaluate RPHunter on a manually-labeled dataset of 645 Rug Pull tokens and an open-source dataset of 1675 benign tokens. The results show that RPHunter effectively detects Rug Pull tokens on the dataset, outperforming five state-of-the-art methods with an acceptable time overhead. Moreover, we apply RPHunter on the Ethereum Mainnet. It has identified 4801 Rug Pull tokens achieving a precision of 90.7\%, which demonstrates RPHunter's effectiveness in real-world scenarios.

\bibliographystyle{IEEEtran}
\bibliography{RPhunter}

\vfill

\end{document}